\documentclass{llncs}

\usepackage[final]{microtype}
\usepackage{lmodern}
\usepackage{amsmath,amssymb,xcolor,graphicx,wrapfig,paralist}
\usepackage{booktabs}
\usepackage{proof}
\usepackage{tikz}
\usepackage{csquotes}
\usepackage{graphics}
\usepackage{stmaryrd}
\usepackage[caption=false]{subfig}
\usepackage{multirow}
\usepackage{makecell}
\usepackage{forest}
\usepackage{pgfplots}
\usepackage{pgfplotstable}
\usepackage{ellipsis, mparhack, ragged2e} 
\usepackage[l2tabu, orthodox]{nag} 

\usetikzlibrary{shapes,positioning,arrows,calc,automata,matrix,fit}
\tikzstyle{state}+=[minimum size = 6mm, inner sep=0,outer sep=1]
\tikzset{->,>=stealth'}
\usetikzlibrary{decorations.markings,decorations.pathmorphing}
\tikzstyle{node1}+=[circle,draw,minimum size = 4mm, inner sep=0,outer sep=0,fill=black]
\tikzstyle{node2}+=[circle,draw,minimum size = 4mm, inner sep=0,outer sep=0,fill=black!50]
\tikzstyle{leaf}+=[circle,draw,minimum size = 4mm, inner sep=0,outer sep=0]
\tikzstyle{select}+=[fill=red]
\tikzstyle{expand}+=[draw=red]
\tikzstyle{rollout}+=[draw=red,thick]

\usepackage{algorithm,algorithmicx} 
\usepackage[noend]{algpseudocode} 
\algnewcommand\algorithmicinput{\textbf{\textsc{Input}:} }
\algnewcommand\Input{\State \algorithmicinput}
\algnewcommand\algorithmicoutput{\textbf{\textsc{Output}:} }
\algnewcommand\Output{\State \algorithmicoutput}
\algdef{SE}[DOWHILE]{Do}{doWhile}{\algorithmicdo}[1]{\algorithmicwhile\ #1}
\algdef{S}[FOR]{ForEach}[1]{\algorithmicfor\ #1\ \algorithmicdo}%

\usepackage{tabu,multirow}
\newcolumntype{L}{X}
\newcolumntype{R}{>{\raggedleft\arraybackslash}X}
\newcolumntype{C}{>{\centering\arraybackslash}X}

\input{macros}

\newcommand{\todojan}[1]{\todo{\textbf{J:\\}#1}}

\renewcommand{\todo}[1]{}

\title{Monte Carlo Tree Search for Verifying Reachability in Markov Decision Processes\thanks{This research was supported in part by
		Deutsche Forschungsgemeinschaft  (DFG)  through  the  TUM  International  Graduate  School  of  Science  and Engineering (IGSSE) project 10.06 \emph{PARSEC}, 
		the Czech Science Foundation grant No.~\mbox{18-11193S}, 
		and 
		the DFG project 383882557 \emph{Statistical Unbounded Verification}.
}}
\author{
	Pranav Ashok\inst{1}, 
	Tom\'{a}\v{s} Br\'{a}zdil\inst{2}, 
	Jan K{\v r}et\'insk\'y\inst{1}
	and Ond{\v r}ej Sl\'{a}me{\v c}ka\inst{2}
}
\institute{Technical University of Munich, Germany \and
	Masaryk University, Brno, Czech Republic}

\begin{document}

\maketitle

\begin{abstract}
	The maximum reachability probabilities in a Markov decision process can be computed using value iteration (VI).
	Recently, simulation-based heuristic extensions of VI have been introduced, such as bounded real-time dynamic programming (BRTDP), which often manage to avoid explicit analysis of the whole state space while preserving guarantees on the computed result.
	In this paper, we introduce a new class of such heuristics, based on Monte Carlo tree search (MCTS), a technique celebrated in various machine-learning settings.
	We provide a spectrum of algorithms ranging from MCTS to BRTDP.
	We evaluate these techniques and show that for larger examples, where VI is no more applicable, our techniques are more broadly applicable than BRTDP with only a minor additional overhead.
\end{abstract} 

\section{Introduction} \label{sec:intro}

{Markov decision processes} (MDP) \cite{Puterman} are a classical formalism for modelling systems with both non-deterministic and probabilistic behaviour.
Although there are various analysis techniques for MDP that run in polynomial time and return precise results, such as those based on linear programming, they are rarely used.
Indeed, dynamic programming techniques, such as value iteration (VI) or policy iteration, are usually preferred despite their exponential complexity.
The reason is that for systems of sizes appearing in practice not even polynomial algorithms are useful and heuristics utilizing the structure of the human-designed systems become a necessity.
Consequently, probabilistic model checking has adopted \cite{atva,cav17,atva17,maxi,storm} techniques, which generally come with weaker guarantees on correctness and running time, but in practice perform better. 
These techniques originate from reinforcement learning, such as delayed Q-learning \cite{pac}, or probabilistic planning, such as bounded real-time dynamic programming (BRTDP) \cite{BRTDP}.

Since verification techniques are applied in safety-critical settings, the results produced by the techniques are required to be correct and optimal, or at least $\varepsilon$-optimal for a given precision $\varepsilon$.
To this end, we follow the general scheme of \cite{atva} and combine the non-guaranteed heuristics with the traditional guaranteed techniques such as VI.
However, while pure VI analyses the whole state space, the heuristics often allow us to focus only on a small part of it and still give precise estimates of the induced error.
This approach was applied in \cite{atva}, yielding a variant of BRTDP for MDP with reachability.
Although BRTDP has already been shown to be quite good at avoiding unimportant parts of the state space in many cases, it struggles in many other settings, for instance where the paths to the goal are less probable or when the degree of non-determinism is high.

In this paper, we go one step further and bring a yet less guaranteed, but yet more celebrated technique of \emph{Monte Carlo tree search} (MCTS) \cite{AK87,Kocsis06banditbased,remi,mctssurvey} into verification.
MCTS is a heuristic search algorithm which combines exact computation using search trees with sampling methods.
To find the best actions to perform, MCTS constructs a search tree by successively unfolding the state-space of the~MDP. The value of each newly added state is evaluated using simulations (also called roll-outs) and its value is backpropagated through the already existing search tree. 

We show that the exact construction of the search tree in MCTS mitigates some of the pitfalls of BRTDP which relies completely on simulation. Namely, the search tree typically reaches less probable paths much sooner than a BRTDP simulation, e.g., in the example depicted in Fig.~\ref{Ex:MCTS-best}.
We combine MCTS with BRTDP in various ways, obtaining thus a spectrum of algorithms ranging from pure BRTDP to pure MCTS along with a few hybrids in between. The aim is to overcome the weaknesses of BRTDP, while at the same time allowing to tackle large state spaces, which VI is unable to handle, with guaranteed ($\varepsilon$)-optimality of the solution. 

While usually performing comparable to BRTDP, we are able to provide reasonable examples which can be tackled using neither BRTDP nor VI, but with our MCTS-BRTDP hybrid algorithms.
Consequently, we obtain a technique applicable to larger systems, unlike VI, which is more broadly applicable than BRTDP with not much additional overhead.

Our contribution can be summarized as follows:
\begin{itemize}
	\item We provide several ways of integrating MCTS into verification approaches so that the resulting technique is an anytime algorithm, returning the maximum reachability probability in MDP together with the respective error bound.
	\item We evaluate the new techniques and compare them to the state-of-the-art implementations based on VI and BRTDP. 
	We conclude that for larger systems, where VI is not applicable, MCTS-based techniques are more robust than BRTDP.
\end{itemize}

\subsection{Related Work}

The correctness of the error bounds in our approach is guaranteed through the computation of the lower and upper bounds on the actual value.
Such a computation has been established for MDP in \cite{atva,II} and the technique is based on the classical notion of the MEC quotient \cite{DeAlfaro1997}.


\paragraph{Statistical model checking (SMC)} \cite{younes02} is a collection of simulation-based techniques for verification where confidence intervals and probably approximately correct results (PAC) are sufficient.
While on Markov chain it is essentially the Monte Carlo technique, on MDP it is more complex and its use is limited \cite{isola-SMCsurvey}, resulting in either slow \cite{atva} or non-guaranteed \cite{qest12,lightweight} methods.
In contrast, MCTS combines Monte Carlo evaluation with explicit analysis of parts of the state space and thus opens new ways of integrating simulations into MDP verification.
	

\paragraph{Monte Carlo tree search (MCTS)} There is a huge amount of literature on various versions of MCTS and applications. See~\cite{mctssurvey} for an extensive survey.
MCTS has been spectacularly successful in several domains, notably in playing classical board games such as Go~\cite{alphago} and chess~\cite{DBLP:journals/corr/abs-1712-01815}.

Many variants of MCTS can be distinguished based on concrete implementations of its four phases: Selection and expansion, where a search tree is extended, roll-out, where simulations are used to evaluate newly added nodes, and back-up, where the result of roll-out is propagated through the search tree.
In the selection phase, actions can be chosen based on various heuristics such as the most common UCT~\cite{Kocsis06banditbased} and its extensions such as FPU and AMAF~\cite{gelly:hal-00115330}. The evaluation phase has also been approached from many directions. To improve upon purely random simulation, domain specific rules have been employed to guide the choice of actions~\cite{ST09} and (deep) reinforcement learning (RL) techniques  have been used to learn smart simulation strategies~\cite{alphago}. On the other hand, empirical evidence shows that it is often more beneficial to make simple random roll-outs as opposed to complex simulations strategies~\cite{JamesKR17}.
RL has also been integrated with MCTS to generalize UCT by temporal difference (TD) backups~\cite{SSM12,VodopivecSS17}.

%


\section{Preliminaries} \label{sec:prelims}  

\subsection{Markov decision processes}
A \emph{probability distribution} on a finite set $X$ is a mapping $d: X\mapsto [0,1]$, such that $\sum_{x\in X} d(x) = 1$. We denote by $\Distributions(X)$ the set of all probability distributions on $X$.
Further, the \emph{support} of a probability distribution $\rho$ is denoted by $\supp(d) = \{x \in X \mid d(x)> 0 \}$.
\begin{definition}[MDP]
	A \emph{Markov decision processes (MDP)} is a tuple of the form $\Mdp = (\states, \initstate, \actions, \av, \trans,\mathfrak 1,\mathfrak 0)$, where $\states$ is a finite set of \emph{states}, $\initstate,\mathfrak 1,\mathfrak 0 \in \states$ is the \emph{initial} state, \emph{goal} state, and \emph{sink} state, respectively, $\actions$ is a finite set of \emph{actions}, $\av: \states \to 2^{\actions}$ assigns to every state a set of \emph{available} actions, and $\trans: \states \times \actions \to \distributions(\states)$ is a \emph{transition function} that given a state $s$ and an action $a\in \av(s)$ yields a probability distribution over successor states.
\end{definition}
An \emph{infinite path} $\path$ in an MDP is an infinite word $\path = s_0 a_0 s_1 a_1 \dots \in (\states \times \actions)^\omega$, such that for every $i \in \Naturals$, $a_i\in \av(s_i)$ and $\trans(s_i,a_i, s_{i+1}) > 0$.
A \emph{finite path} $\fpath = s_0 a_0 s_1 a_1 \dots s_n \in (\states \times \actions)^* \times \states$ is a finite prefix of an
infinite path.
A \emph{policy} on an MDP is a function $\straa: (\states \times \actions)^*\times S \to \distributions(\actions)$, which given a finite path $\fpath = s_0 a_0 s_1 a_1 \dots s_n$ yields a probability distribution $\straa(\fpath) \in \distributions(\av(s_n))$ on the actions to be taken next.
We denote the set of all strategies of an MDP by $\straas$.
Fixing a policy $\straa$ and an initial state $s$ on an MDP $\Mdp$ yields a unique probability measure $\pr^\straa_{\Mdp, s}$ over infinite paths~\cite[Sect.~2.1.6]{Puterman} and thus the probability $V^\straa(s):=\PrS{\straa}{\Mdp, s}{\{ \rho \mid \exists i \in \Naturals : \rho(i) =\mathfrak 1\}}$ to reach the goal state when following $\straa$.


\begin{definition}
	Given a state $s$, the \emph{maximum reachability probability} is 
	$V(s) = \sup_{\straa \in \straas}  V^\straa(s)$.
	A policy $\sigma$ is $\varepsilon$-optimal if $V(\initstate) - V^{\sigma}(\initstate) \leq \varepsilon$.
\end{definition}
Note that there always exists a 0-optimal policy of the form $\straa: \states \to \actions$ \cite{Puterman}.	

A pair $(T, A)$, where $\emptyset \neq T \subseteq S$ and $\emptyset \neq A \subseteq \Union_{s\in T} \av(s)$, is an \emph{end component} of an MDP $\Mdp$ if 
\begin{itemize}
	\item for all $s \in T, a \in A \intersection \av(s)$ we have $\supp(\trans(s,a)) \subseteq T$, and 
	\item for all $s, s' \in T$ there is a finite path $\fpath = s a_0 \dots a_n s' \in (T \times A)^* \times T$, i.e.\ $\fpath$ starts in $s$, ends in $s'$, stays inside $T$ and only uses actions in $A$.
\end{itemize}
An end component $(T, A)$ is a \emph{maximal end component (MEC)} if there is no other end component $(T', A')$ such that $T \subseteq T'$ and $A \subseteq A'$.
The set of MECs of an MDP $\Mdp$ is denoted by $\mec(\Mdp)$.
An MDP can be turned into its \emph{MEC quotient}~\cite{DeAlfaro1997} as follows. 
Each MEC is merged into a single state, all the outgoing actions are preserved, except for a possible self-loop (when all its successors are in this MEC). 
Moreover, if there are no available actions left then it is merged with the sink state $\mathfrak 0$. For a formal definition, see Appendix~\ref{app:quotient}.
The techniques we discuss in this paper require the MDP be turned (gradually or at once) into its MEC-quotient in order to converge to $V$, as described below.

\subsection{Value Iteration}

Value iteration (VI) is a dynamic programming algorithm first described by Bellman \cite{bellman}. The maximum reachability probability is characterized as the least fixpoint of the following equation system for $s\in\states$:
\[
V(s) =
\begin{cases}
	1 & \text{if }s =\mathfrak 1 \\
	0 & \text{if }s =\mathfrak 0\\
	\max\limits_{a \in Av(s)} \sum\limits_{s' \in S} \Delta(s,a)(s') \cdot V(s')
	& \text{otherwise} 
\end{cases}
\]

%

In order to compute the least fixpoint, an iterative method can be used as follows.
We initialize $V^0(s) = 0$ for all $s$ except for $V^0(\mathfrak 1)=1$.
A successive iteration $V^{i+1}$ is computed by evaluating the right-hand side of the equation system with $V^i$.
The optimal value is achieved in the limit, i.e. $\lim_{n \to \infty} V^n=V$. 

%
%

In order to obtain bounds on the imprecision of $V^n$, one can employ a \emph{bounded} variant of VI \cite{BRTDP,atva} (also called \emph{interval iteration} \cite{II}).
Here one computes not only the \emph{lower bounds} on $V$ via the least-fixpoint approximants $V^n$ (onwards denoted by $L^n$), but also \emph{upper bounds} via the greatest-fixpoint approximants, called $U^n$.
The greatest-fixpoint can be approximated in a dual way, by initializing the values to $1$ except for $\mathfrak 0$ where it is $0$.
On MDPs without MECs (except for $\mathfrak 1$ and $\mathfrak 0$), we have both $\lim_{n\to\infty}L^n=V=\lim_{n\to\infty} U^n$, giving us an anytime algorithm with the guarantee that $V\in[L^n,U^n]$.
However, on general MDPs, $\lim_{n\to\infty} U^n$ can be larger than $V$, often yielding only a trivial bound of $1$, see Appendix~\ref{app:U-fail}.
The solution suggested in \cite{atva,II} is to consider the MEC quotient instead, where the guarantee is recovered.
The MEC quotient can be pre-computed \cite{II} or computed gradually on-the-fly only for a part of the state space \cite{atva}.


\subsection{Bounded Real-Time Dynamic Programming (BRTDP)}
\label{sec:brtdp}

BRTDP \cite{BRTDP} is a heuristic built on top of (bounded) VI, which has been adapted to the reachability objective in \cite{atva}.
It belongs to a class of \textit{asynchronous} VI algorithms.
In other words, it differs from VI in that in each iteration it does not update the values for all states, but only a few.
The states to be updated are chosen as those that appear in a simulation run performed in that iteration. 
This way we focus on states visited with high probability and thus having higher impact on the value.

BRTDP thus proceeds as follows.
It maintains the current lower and upper bounds $L$ and $U$ on the value, like bounded VI.
In each iteration, a simulation run is generated by choosing in each state $s$
\begin{itemize}
	\item the action maximizing $U$, i.e. $\arg\max_a \sum_{s'\in\states}\trans(s,a)(s')\cdot U(s')$,
	\item the successor $s'$ of $a$ randomly with weight $\trans(s,a)(s')$ or (more efficiently) in the variant of \cite{atva} with weight $\trans(s,a)(s') \cdot (U(s') - L(s'))$.
\end{itemize}
Then the value of each state on the simulation run is updated by the equation system for $V$.
This happens preferably in the backward order \cite{atva} from the last state ($\mathfrak 1$ or $\mathfrak 0$) towards $\initstate$, thus efficiently propagating the value information. 
These iterations are performed repeatedly until $U(\initstate)-L(\initstate)<\varepsilon$ for some predefined precision $\varepsilon$, yielding the interval $[L(\initstate),U(\initstate)]$ containing the value $V(\initstate)$. 

If the only MECs are formed by $\mathfrak 1$ and $\mathfrak 0$ then the algorithm (depicted in Appendix~\ref{app:brtdp} as Algorithm \ref{alg:brtdp}) terminates almost surely.
But if there are other MECs, then this is not necessarily the case.
In order to ensure correct termination, \cite{atva} modifies the algorithm in a way, which we adopt also for our algorithms: 
Periodically, after a certain number of iterations, we compute the MEC quotient, not necessarily of the whole MDP, but only of the part formed by the states visited so far by the algorithm.


\section{Monte-Carlo Tree Search} \label{sec:mcts}

\subsubsection*{Motivation}

\begin{figure}[t]
	\begin{center}
		\begin{tikzpicture}
		\tikzstyle{state}=[thick,draw=black,circle]
		\tikzstyle{transition}=[draw,shape=circle,fill=black]
		\tikzstyle{loop}=[looseness=5, in=120, out=60]
		
		\node[state] at (0,0) (s0) {0};
		\node[transition] at (1.6,0) (s0b) {};
		\node[state] at (3.2,0) (s1) {1};
		\node[transition] at (4.8,0) (s1b) {};
		\node[state] at (6.4,0) (s2) {2};
		\node[transition] at (8,0) (s2b) {};
		\node[state] at (9.6,0) (s3) {3};
		
		\draw (s0) edge [bend left, ->] node [midway, above] {$a$} (s0b);
		\draw[->] (s0b) -- (s1) node [midway, above] {0.01};
		\draw (s0b) edge [bend left, ->] node [midway, above] {0.99} (s0);
		
		\draw (s1) edge [bend left, ->] node [midway, above] {$b$} (s1b);
		\draw[->] (s1b) -- (s2) node [midway, above] {0.01};
		\draw (s1b) edge [bend left, ->] node [midway, above] {0.99} (s0);
		
		\draw (s2) edge [bend left, ->] node [midway, above] {$c$} (s2b);
		\draw[->] (s2b) -- (s3) node [midway, above] {0.01};
		\draw (s2b) edge [bend left, ->] node [midway, above] {0.99} (s0);
		
		\draw (s3) edge [loop,->] node [above] {$d$} (s3);
		\end{tikzpicture}
	\end{center}
	\caption{A difficult case for BRTDP}
	\label{fig:worstcase}
\end{figure}
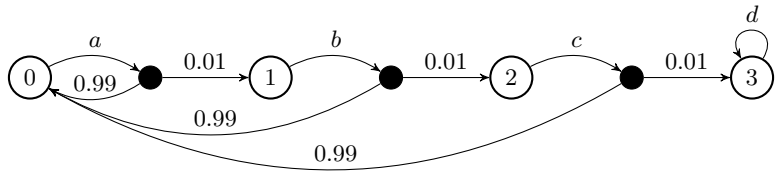

One of the main challenges in the application of BTRDP are the presence of events that happen after a long time (but are not necessarily rare). 
For example, consider the simple MDP in Fig. \ref{fig:worstcase}, modelling a hypothetical communication protocol. 
Let $s_3$ be the goal state, representing a failure state. 
From each of the states $s_0, s_1$ and $s_2$, the MDP tries to send a message. 
If the message sending fails, which happens with a very low probability, the MDP moves to the next state, otherwise returning to the initial state $s_0$. 
BRTDP repeatedly samples paths and propagates (``backs-up'') the encountered values through the paths. 
Even though the goal is reached in almost every simulation, it can take very long time to finish each simulation.

The idea of MCTS is to ``grow'' a tree into the model rooted at the initial state while starting new simulations from the leaves of the tree. In this example, the tree soon expands to $s_1$ or $s_2$ and from then on, simulations are started there and thus are more targeted and the  back-propagation of the values is faster.
 

\subsubsection*{Generic algorithm}

\begin{figure}[t]
	\centering
	\minipage[t]{0.25\textwidth}
	\vspace{0pt}
	\begin{forest}
		[,node1,select
		[,node1,edge={-latex,thick},select
		[,leaf,edge={-}]
		[,leaf,edge={-latex,thick},select]
		]
		[,node1,edge={-}]
		[,node1,edge={-}
		[,leaf,edge={-}]
		[,leaf,edge={-}]
		]
		]
	\end{forest}
	\endminipage \hfill
	\minipage[t]{0.25\textwidth}
	\vspace{0pt}
	\begin{forest}
		[,node1,edge={-}
		[,node1,edge={-}
		[,leaf,edge={-}]
		[,node1,edge={-}
		[,leaf,expand,edge=dotted]
		[,leaf,expand,edge=dotted]
		[,leaf,expand,edge=dotted]
		]
		]
		[,node1,edge={-}]
		[,node1,edge={-}
		[,leaf,edge={-}]
		[,leaf,edge={-}]
		]
		]
	\end{forest}
	\endminipage \hfill
	\minipage[t]{0.25\textwidth}
	\vspace{0pt}
	\begin{forest}
		[,node1,edge={-}
		[,node1,edge={-}
		[,leaf,edge={-}]
		[,node1,edge={-}
		[,leaf,edge={-}]
		[,leaf,edge={-}]
		[,leaf,rollout,edge={-}
		[,edge={-latex, thick, decorate, decoration={snake,amplitude=0.5mm,post length=4pt}, draw=red},l*=1.6]
		]
		]
		]
		[,node1,edge={-}]
		[,node1,edge={-}
		[,leaf,edge={-}]
		[,leaf,edge={-}]
		]
		]
	\end{forest}
	\endminipage \hfill
	\minipage[t]{0.25\textwidth}
	\vspace{0pt}
	\begin{forest}
		[,node2,edge={-}
		[,node2,edge={latex-,thick}
		[,leaf,edge={-}]
		[,node2,edge={latex-,thick}
		[,leaf,edge={-}]
		[,leaf,edge={-}]
		[,leaf,rollout,edge={latex-,thick},draw=red
		[,edge+={latex-,thick,decorate, decoration={snake,amplitude=0.5mm,pre length=6pt}, draw=red}, l*=1.6]
		]
		]
		]
		[,node1,edge={-}]
		[,node1,edge={-}
		[,leaf,edge={-}]
		[,leaf,edge={-}]
		]
		]
	\end{forest}
	\endminipage
	\caption{The four stages of MCTS: \textit{selection}, \textit{expansion}, \textit{roll-out} and \textit{back-propagation}}
	\label{fig:mcts}
\end{figure}
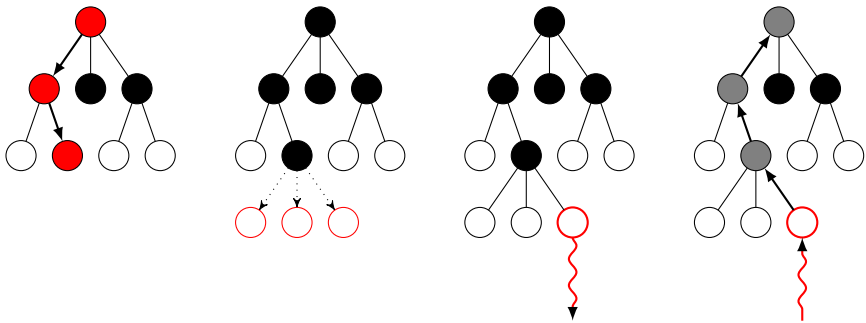

MCTS can be visualized (see Fig. \ref{fig:mcts}) as gradually building up a tree, which describes several steps of unfolding of the system, collecting more and more information. {
Note that nodes of the tree correspond to states of the MDP.}

The MCTS procedure (see Algorithm \ref{algo:mcts}) proceeds in rounds, each consisting of several stages:

First, in the \textbf{selection} stage (line \ref{l:select}), a \textit{tree policy} is followed leading from the root of the tree to the currently most ``interesting'' leaf.

Second, in the \textbf{expansion} stage (line \ref{l:expand}), one or more of the successors are added to the tree. 
In our setting with required guarantees, it turns out sensible to expand all successors, i.e. $\bigcup_{a\in\av(t_{\mathrm{parent}})} \supp(\trans(s_{\mathrm{parent}},a))$.
One of the successors is picked, according to the tree policy.

Third, in the \textbf{roll-out} stage (line \ref{l:roll-out}) a simulation run is generated, adhering to a \textit{roll-out policy}.
In the classical setting, the simulation run receives a certain reward. 
In our setting, the simulation run receives a reward of 1 if it encounters the goal state, otherwise it receives 0.

Finally, in the \textbf{back-propagation} (or \textbf{backup}) stage, the information whether a goal was reached during the roll-out is propagated up the tree (line \ref{l:bp-tree}). 
The information received through such simulations helps MCTS to decide the direction in which the tree must be grown in the next step.
On top of the standard MCTS, we also consider possible back-propagation of this information along the generated simulations run (line \ref{l:bp-run}).

Next, we discuss a particularly popular way to implement the MCTS scheme.

\begin{algorithm}[t]
	\caption{The (enriched) MCTS scheme}
	\label{algo:mcts}
	\begin{algorithmic}[1]
		\State Create a root $t_0$ of the tree, labelled by $\initstate$
		\While{within computational budget}
		\State $t_{\mathrm{parent}} \gets \Call{SelectNode}{t_0}$ \label{l:select}
		\State $t_{\mathrm{child}} \gets \Call{ExpandAndPickroll-outNode}{t_{\mathrm{parent}}}$ \label{l:expand}
		\State $\mathit{outcome} \gets \Call{roll-out}{t_{\mathrm{child}}}$  \label{l:roll-out}
		\State $\Call{BackupOnRoll-out}{\mathit{outcome}}$ \Comment Not in the standard MCTS \label{l:bp-run}
		\State $\Call{BackupOnTree}{t_{\mathrm{child}}, \mathit{outcome}}$ \label{l:bp-tree}
		\EndWhile
		\State \Return \Call{InducedPolicy}{$t_0$} \Comment{ action with the best estimated value}
	\end{algorithmic}
\end{algorithm}

\subsubsection*{UCT Implementation}

Each stage of the MCTS scheme can be instantiated in various ways. 
We now describe one of the most common and successful implementations, called UCT \cite{Kocsis06banditbased} (abbreviation of ``UCB applied to Trees'').
Each node $t$ in the tree keeps track of two numbers: $n_t$ is the number of times $t$ has been visited when selecting leaves;
$v_t$ is the number of times it has been visited \emph{and} the roll-out from the descendant leaf hit the goal.
These numbers can be easily updated during the back-propagation: we simply increase all $n$'s on the way from the leaf to the root, similarly for the $v$'s whenever the roll-out hit the goal.
This allows us to define the last missing piece, namely the tree policy based on these values, which is called \textit{Upper Confidence Bound} (UCB1 or simply UCB). 
The UCB1 value of a node $t$ in the MCTS tree is defined as follows:
\begin{equation*}
\mathit{UCB1}(t) = \frac{v_t}{n_t} + C \sqrt{ \frac{ \ln n_{\mathrm{parent}(t)}}{n_t} }
\end{equation*}
where $\mathrm{parent}(t)$ is the parent of $t$ in the tree and $C$ is a fixed \emph{exploration-exploitation constant}.
{
The tree policy choosing nodes with maximum UCB1 bound is called the {\em UCB1 policy}.}

Intuitively, the first term of the expression describes the ratio of ``successful'' roll-outs (i.e., hitting the goal) started from a descendant of $t$. 
In other words, this \emph{exploitation} term tries to approximate the probability to reach the goal from $t$.
The higher the value, the higher the chances that the tree policy would pick $t$. 
In contrast, the second term  decreases when the node $t$ is chosen too often. 
This compensates for the first term, thereby allowing for the siblings of $t$ to be explored.
This is the \emph{exploration} term and the effect of the two terms is balanced by the constant $C$. 
The appropriate values of $C$ vary among domains and, in practice, are determined experimentally. 
 

It has been proved in~\cite{Kocsis06banditbased} that for finite-horizon MDP (or MDP with discounted rewards), the UCT with an appropriately selected $C$ converges in the following sense:
After $n$ iterations of the MCTS algorithm, the difference between the expectation of the value estimate and the true value of a given state is bounded by $\mathcal{O}(\log(n)/n)$. Moreover, the probability that an action with the best upper confidence bound in the root is not optimal converges to zero at a polynomial rate with $n\rightarrow \infty$.

\section{Augmenting MCTS with guarantees}

In this section,  we present several algorithms based on MCTS.
Note that the typical uses of MCTS in practice only require the algorithm to guess a good but not necessarily the optimal action for the current state. 
When adapting learning algorithms to the verification setting, an important requirement is an ability to give precise guarantees on the sought value or to produce an $\varepsilon$-optimal scheduler.
Consequently, in order to obtain the guarantees, our algorithms combine MCTS with BRTDP, which comes with guarantees, to various degrees.
The spectrum is depicted in Table \ref{tab:spectrum} and described in more detail below.

\begin{table}[]
	\centering
	\caption{Spectrum of algorithms ranging from pure MCTS to pure BRTDP}
	\label{tab:spectrum}
	\begin{tabular}{llllll}
		\toprule
		\thead{Algorithm}                                     & \thead{MCTS}~~~ & \thead{BMCTS}~~~ & \thead{MCTS-\\BRTDP}~~~ & \thead{BRTDP-\\UCB}~~~        & \thead{BRTDP}~~~         \\ \midrule
		\makecell[r]{\textsc{SelectNode} (Tree Policy)}~~~ & UCB1            & UCB1             & UCB1                    & \multirow{2}{*}{UCB1}         & \multirow{2}{*}{BRTDP}   \\
		\makecell[r]{\textsc{Roll-out}}~~~                    & Uniform         & Uniform          & BRTDP                   &                               &  \\
		\makecell[r]{\textsc{BackpupOnRoll-out}}~~~           & ---             & $L,U$            & $L,U$                   & \multirow{2}{*}{$v,n,\  L,U$} & \multirow{2}{*}{$L,U$}   \\
		\makecell[r]{\textsc{BackupOnTree}}~~~                & $v,n$           & $v, n,\ L,U$     & $v,n,\ L,U$             &                               &  \\ \bottomrule
	\end{tabular}
\end{table}

\begin{enumerate}
	\item Pure MCTS (the UCT variant): The tree is constructed using the UCB1 policy while roll-outs are performed using a uniform random policy, i.e. actions are chosen uniformly, successors according to their transition probabilities. The roll-outs are either successful or not, and this information is back-propagated up the tree, i.e. the $v$- and $n$-values of the nodes in the tree are updated.
	\item Bounded MCTS (BMCTS): In addition to all the steps of pure MCTS, here we also update the $L$- and $U$-values, using the transition probabilities and the old $L$- and $U$-values.
	This update takes place both on all states in the tree on the path from the root to the current leaf as well as all states of visited during the roll-out.
	The updates happen backwards, starting from the last state of the simulation run, towards the leaf and then towards the root.
	\item MCTS-BRTDP: This algorithm is essentially BMCTS with the only difference that the roll-out is performed using the BRTDP policy, i.e. action is chosen as $\arg\max_{a\in\av(s)}\sum_{s'}\trans(s,a)(s')\cdot U(s')$, and the successor of $a$ randomly with the weight $\trans(s,a)(s') \cdot (U(s') - L(s'))$.
	\item BRTDP-UCB: As we move towards the BRTDP side of the spectrum, there is no difference between whether a state is captured in the tree or not: back-propagation works the same and the policy to select the node is the same as for the roll-out.
	In this method, we use the UCB1 policy to choose actions on the whole path from the initial state all the way to the goal or the sink, and back-propagate all information.
	Note that as opposed to BRTDP,  the exploitation is interleaved with some additional exploration, due to UCB1.
	\item Pure BRTDP: Finally, we also consider the pure BRTDP algorithm as presented earlier in the paper.
	This works the same as BRTDP-UCB, but the (selection and roll-out) policy is that of BRTDP.
\end{enumerate}

While MCTS does not provide exact guarantees on the current error, all the other methods keep track of the lower and the upper bound, yielding the respective guarantee on the value.
Note that MCTS-BRTDP is a variant of MCTS, where BRTDP is used not only to provide these bounds, but also to provide a more informed roll-out.
Such a policy is expected to be more efficient compared to just using a uniform policy or UCB. Since some studies \cite{JamesKR17} have counter-intuitively shown that more informed roll-outs do not necessarily improve the runtime, we also include BMCTS and BRTDP-UCB, where the path generation is derived from the traditional MCTS approach; the former applies it in the MCTS setting, the latter in the BRTDP setting.

\subsubsection*{MDPs with MECs}

In MDPs where the only MECs are formed by $\mathfrak 1$ and $\mathfrak 0$, the roll-outs almost surely reach one of the two states and then stop.
Since in MDPs with MECs this is not necessarily the case, we have to collapse the MECs, like discussed in Section~\ref{sec:brtdp}.
Therefore, a roll-out $w = s_0 s_1 s_2 \dots s_n$ is stopped if $s_n\in\{\mathfrak 1,\mathfrak 0\}$ or $s_n = s_k$, for some $0 \leq k < n$.
In the latter case, there is a chance that an end component has been discovered.
Hence we compute the MEC quotient and only then continue the process.
When MECs are collapsed this way, both the lower and the upper bound converge to the value and the resulting methods terminate in finite time almost surely, due to reasoning of \cite{atva} and the exploration term being positive. \todojan{very handwavy, we should go through the argument, but maybe only for the final version\\J: not now}

\begin{example}\label{Ex:MCTS-best}
We revisit the example of Fig.~\ref{fig:worstcase} to see how the MCTS-based algorithms presented above tend to work. 
We focus on MCTS-BRTDP. 
Once the algorithm starts it is easy to see that in 3 iterations of MCTS-BRTDP, the target state $s_3$ will belong to the MCTS tree. From the next iteration onwards, the selection step will always choose $s_3$ to start the roll-out from. But since $s_3$ is already a target, the algorithm just proceeds to update the information up the tree to the root state. Hence, with just 3 iterations more than what value iteration would need, the algorithm converges to the same result.

While this example is quite trivial, we show experimentally in the next section that the MCTS-based algorithms not only run roughly as fast as BRTDP, but in certain large models that exhibit behaviour similar to the example of Fig. \ref{fig:worstcase}, it is clearly the better option.
\end{example}



\section{Experimental Evaluation} \label{sec:experiments}

We implemented all the algorithms outlined in Table \ref{tab:spectrum} in PRISM Model Checker \cite{prism}. Table \ref{tab:main} presents the run-times obtained on twelve different models. 
Six of the models used (\texttt{coin, leader, mer, firewire, zeroconf} and \texttt{wlan}) were chosen from the PRISM Benchmark Suite \cite{prismbenchmarks} and the remaining six were constructed by modifying \texttt{firewire}, \texttt{zeroconf} and \texttt{wlan} in order to demonstrate the strengths and weaknesses of VI, BRTDP, and our algorithms. 
Recall the motivational example of Fig. \ref{fig:worstcase} in Section \ref{sec:mcts}, which is hard for BRTDP, yet easy for VI. We refer to this MDP as \texttt{brtdp-adversary}.
This hard case is combined with one of benchmarks, \texttt{firewire}, \texttt{zeroconf} and \texttt{wlan}, in two ways as follows.

\begin{figure}[t]
	\begin{center}
		\begin{tikzpicture}
		\tikzstyle{state}=[thick,draw=black,circle]
		\tikzstyle{transition}=[draw,shape=circle,fill=black]
		\tikzstyle{loop}=[looseness=5, in=120, out=60]
		
		\node[state] at (0,0) (s0) {$s_0$};
		\node[transition] at (1.6,0) (s0b) {};
		\node[state] at (3.2,0) (s1) {$s_1$};
		\node[transition] at (4.8,0) (s1b) {};
		\node[state] at (6.4,0) (s2) {$s_2$};
		\node[transition] at (8,0) (s2b) {};
		\node[state] at (9.6,0) (s3) {$s_3$};
		
		\node [cloud, draw,cloud puffs=10,cloud puff arc=120, aspect=2,
		inner ysep=1em] at (2,-2.5) (zconf) {zeroconf};
		
		\draw (s0) edge [bend right, ->] node [midway, right] {$b$} (zconf);
		
		\draw (s0) edge [bend left, ->] node [midway, above] {$a$} (s0b);
		\draw[->] (s0b) -- (s1) node [midway, above] {0.01};
		\draw (s0b) edge [bend left, ->] node [midway, above] {0.99} (s0);
		
		\draw (s1) edge [bend left, ->] node [midway, above] {} (s1b);
		\draw[->] (s1b) -- (s2) node [midway, above] {0.01};
		\draw (s1b) edge [bend left, ->] node [midway, above] {0.99} (s0);
		
		\draw (s2) edge [bend left, ->] node [midway, above] {} (s2b);
		\draw[->] (s2b) -- (s3) node [midway, above] {0.01};
		\draw (s2b) edge [bend left, ->] node [midway, above] {0.99} (s0);
		
		\draw (s3) edge [loop,->] node [above] {} (s3);
		
		\end{tikzpicture}
	\end{center}
	\caption{Combining zeroconf model with the BRTDP adversary in a
		branching manner.}
	\label{fig:branching}
\end{figure}
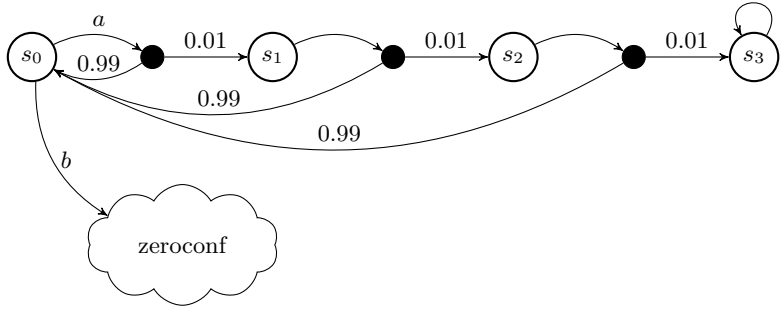

\paragraph{Branch Models} In the first construction, we allow a branching decision at the initial state. The resulting model is shown in Fig. \ref{fig:branching}. If action $a$ is chosen, then we enter the \texttt{brtdp-adversary} branch. If action $b$ is chosen, then we enter the standard \texttt{zeroconf} model. Using this schema, we create three models: \texttt{branch-zeroconf}, \texttt{branch-firewire} and \texttt{branch-wlan}.

\paragraph{Composition Models} In the second construction, the models are constructed by the parallel (asynchronous) composition of the \texttt{brtdp-adversary} and one of \texttt{firewire}, \texttt{zeroconf} and \texttt{wlan}. The resulting models are named \texttt{comp-firewire}, \texttt{comp-zeroconf} and \texttt{comp-wlan}.

\medskip

\begin{table}[h]
	\centering
	\caption{Comparison of the various algorithms. All run-times are in seconds. Cells with `$-$' denote either running out of memory (in the case of VI) or running out of time.}
	\label{tab:main}
	{\renewcommand{\arraystretch}{1.5}%
	\begin{tabular}{lrrrrr}
		\toprule
		\thead{Benchmark}  & \thead{BMCTS} & ~~~\thead{MCTS-\\BRTDP} & ~~~\thead{BRTDP-\\UCB} & ~~~\thead{BRTDP} & ~~~~~~~~~\thead{VI} \\ \midrule

consensus & 5.55 & 6.48 & 7.47 & 6.15 & 1.13 \\

leader & 18.67 & 15.79 & 16.33 & 15.06 & 8.94 \\

mer & $-$ & 4.79 & $-$ & 3.63 & $-$ \\

firewire & 0.07 & 0.08 & 0.09 & 0.09 & 6.99 \\

wlan & 0.09 & 0.07 & 0.08 & 0.08 & $-$ \\

zeroconf & 0.93 & 0.20 & 0.59 & 0.20 & $-$ \\

comp-firewire & 9.36 & 9.55 & $-$ & $-$ & 20.77 \\

comp-wlan & 2.51 & 2.25 & $-$ & $-$ & $-$ \\

comp-zeroconf & $-$ & 29.55 & $-$ & $-$ & $-$ \\

branch-firewire & 0.09 & 0.09 & 0.02 & 0.09 & 9.33 \\

branch-wlan & 0.10 & 0.08 & 0.09 & 0.07 & $-$ \\

branch-zeroconf & 25.90 & 30.78 & 35.67 & 38.14 & $-$ \\ 
		\bottomrule
	\end{tabular}
	}
\end{table}

We run the experiments on an Intel Xeon server with sufficient memory (default PRISM settings). The implementation is single threaded and each experiment is run 15 times to alleviate the effect of the probabilistic nature of the algorithms. The median run-time for each model configuration is reported in Table \ref{tab:main}. Note that the measured time is the wall-clock time needed to perform the model checking. We do not include the time needed to start  PRISM and to read and parse the input file. An execution finishes successfully once the value of the queried maximal reachability property is known with a guaranteed absolute precision of $10^{-6}$, except for \texttt{comp-zeroconf} and \texttt{branch-zeroconf} where it stops once the value is known with a precision of $10^{-2}$. Timeout is set to 10 minutes.

Table~\ref{tab:main} shows that in the cases where BRTDP performs well, our algorithms perform very similarly, irrespective of the performance of VI.
However, when BRTDP struggles due to the hard case, our MCTS-based algorithms still perform well, even in cases where VI also times out.
In particular, MCTS-BRTDP shows a consistently good performance. 
The less informed variants BMCTS and BRTDP-UCB of MCTS-BRTDP and BRTDP, respectively, perform consistently at most as well their respective more informed variants.
This only confirms the intuitive expectation which, however, has been shown invalid in other settings.

In the experiments, we have used the exploration-exploitation constant $C=25$, which is rather large compared to the typical usage in literature \cite{Kocsis2006ImprovedMS,mctssurvey}.
This significantly supports exploration.
The effect of the constant is illustrated in Fig. \ref{fig:constant}.
We can see that for lower values of $C$, the algorithms perform worse, due to insufficient incentive to explore.

\pgfplotstableread{cvalue.data}
\datatable

\begin{figure}[t]
	\centering
\begin{tikzpicture}
\begin{axis}[
width=0.95\textwidth,
height=9cm,
xlabel={$C$},
ylabel={Time (s)},
xmin=0, xmax=100,
ymin=0, ymax=60,
xtick={0,20,40,60,80,100},
ytick={0,10,20,30,40,50},
legend pos=north east,
ymajorgrids=true,
grid style=dotted,
]

\addplot table [y = mer] from \datatable;
\addlegendentry{mer}
\addplot table [y = comp-zconf] from \datatable;
\addlegendentry{comp-zeroconf}
\addplot table [y = comp-firewire] from \datatable;
\addlegendentry{comp-firewire}
\addplot table [y = coin] from \datatable;
\addlegendentry{coin}
\addplot table [y = leader] from \datatable;
\addlegendentry{leader}

\end{axis}
\end{tikzpicture}
\caption{Dependence of the run-time of MCTS-BRTDP on the exploration-exploitation constant, C}
\label{fig:constant}
\end{figure}
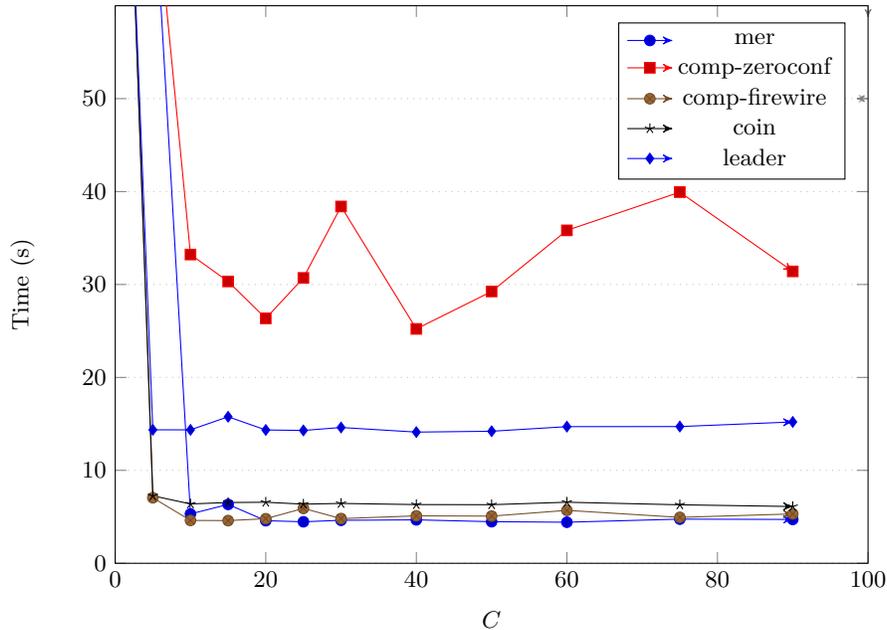

In conclusion, we see that exploration, which is present in MCTS but not explicitly in BRTDP, leads to a small overhead which, however, pays off dramatically in more complex cases.
For the total number of states explored by VI, BRTDP, and our MCTS-based algorithms, we refer the reader to Table \ref{tab:states} in Appendix \ref{app:states}.


\section{Conclusion}

We have introduced Monte Carlo tree search into verification of Markov decision processes, ensuring hard guarantees of the resulting methods.
We presented several algorithms, covering the spectrum between MCTS and (in this context more traditional) MDP-planning heuristic BRTDP.
Our experiments suggest that the overhead caused by additional exploration is outweighed by the ability of the technique to handle the cases, which are more complicated for BRTDP.
Further, similarly to BRTDP, the techniques can handle larger state spaces, where the traditional value iteration fails to deliver any result.
Consequently, the method is more robust and applicable to a larger spectrum of examples, at the cost of a negligible overhead.
\todojan{polish conclusion, also depending on final numbers}

\bibliographystyle{alpha}
\bibliography{ref}

\appendix

\section*{Appendix}

\section{MEC quotient}\label{app:quotient}

A MEC quotient \cite{DeAlfaro1997} is defined as follows:

Let $\Mdp = (\states, \initstate, \actions, \av, \trans,\mathfrak 1,\mathfrak 0)$ be an MDP with MECs $\mec(\Mdp) = \{(T_1, A_1), \dots, (T_n, A_n)\}$.
Further, define $\mec_\states = \Union_{i=1}^n T_i$ as the set of all states contained in some MEC.
The \emph{MEC quotient of} $\Mdp$ is defined as the MDP $\Mdp' = (\states', {s'}_\textrm{init}, \actions', \av', \trans', \mathfrak{1'},\mathfrak{0'})$, where
\begin{itemize}
	\item $\states' = \states \setminus \mec_\states \union \set{{t}_1, \dots, {t}_n}$, where $t_i$ is a representative for $(T_i, A_i)$
	\item if for some $T_i$ we have $\initstate\in T_i$, then ${s'}_\textrm{init} = {t}_i$,  otherwise ${s'}_\textrm{init} = \initstate$,
	\item if for some $T_i$ we have $\mathfrak{1} \in T_i$, then $t_i = \mathfrak{1'}$
	\item if for some $T_i$ we have $\mathfrak{0} \in T_i$, then $t_i = \mathfrak{0'}$
	\item $\actions' =  \set{(s,a) \mid s\in \states, a\in \av(s)}$,
	\item the available actions $\av'$ are defined as
	\begin{align*}
	\forall s \in \states \setminus \mec_\states.~& \av'(s) = \set{(s,a) \mid a \in \av(s)} \\
	\forall 1 \leq i \leq n.~& \av'(t_i) = \set{(s,a) \mid s \in T_i \land a \in \av(s) \setminus A_i},
	\end{align*}
	\item the transition function $\trans'$ is defined as follows.
	Let ${s'} \in {S'}$ be some state in the quotient and $(s, a) \in \av({s'})$ an action available in ${s'}$.
	Then
	\begin{equation*}
	{\trans'}(s', (s, a), s'') = \begin{dcases*}
	{\sum}_{s' \in T_j} \trans(s, a, s') & if $s'' = t_j$, \\
	\trans(s, a, s'') & otherwise, i.e.\ $s'' \in \states \setminus \mec_\states$.
	\end{dcases*}
	\end{equation*}
\end{itemize}

\section{Example of $\lim_{n\to\infty}U^n>V$ due to MECs}
\label{app:U-fail}

\begin{example}
	Below is an MDP with an end component
	$(\{s_0, s_1\}, \{a,c\})$. Let $F = \{s_2\}$.
	When BRTDP updates state $s_0$ or $s_1$ there is always a state (the
	other one in the EC) which has upper bound 1. This way the upper
	bound remains 1 after every iteration, even though the lower bound
	is correctly $0.5$. The algorithm does not converge for $\epsilon <
	0.5$.
	
	
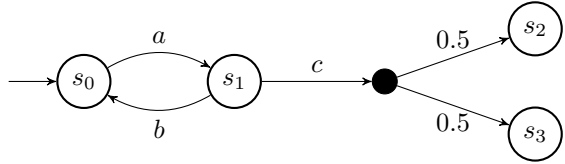
\begin{center}
		\begin{tikzpicture}
		\tikzstyle{state}=[thick,draw=black,circle]
		\tikzstyle{transition}=[draw,shape=circle,fill=black]
		
		\node[state] at (0,0) (s0) {$s_0$};
		
		\draw[<-] (s0) -- node[above] {} ++(-1cm,0);
		
		\node[state] at (2,0) (s1) {$s_1$};
		\node[transition] at (4,0) (s0b) {};
		\node[state] at (6, 0.7) (s2) {$s_2$};
		\node[state] at (6,-0.7) (s3) {$s_3$};
		
		\draw (s0) edge [bend left, ->] node [midway, above] {$a$} (s1);
		\draw (s1) edge [bend left, ->] node [midway, below] {$b$} (s0);
		\draw (s1) edge [->] node [midway, above] {$c$} (s0b);
		
		\draw[->] (s0b) -- (s2) node [midway, above] {0.5};
		\draw[->] (s0b) -- (s3) node [midway, below] {0.5};
		
		\end{tikzpicture}
	\end{center}
\end{example}

\section{BRTDP Algorithm}\label{app:brtdp}

\begin{algorithm}
	\caption{BRTDP for MDPs without end components}
	\label{alg:brtdp}
	\begin{algorithmic}[1]
		\State $U(s,a) \gets 1, L(s,a) \gets 0 \; \forall s \in S, a \in Av(s)$
		\State $U(0,a) \gets 0, L(1,a) \gets 1 \; \forall a \in A$
		\State $\omega \gets s_0$
		\While{$U(s_0) - L(s_0) > \epsilon$ }
		\While{$last(\omega) \not \in \{\mathfrak 0, \mathfrak 1\}$}		\Comment Simulation
		\State $a \gets$ sample uniformly from
		$\argmax\limits_{a \in Av(last(\omega))}
		U(last(\omega), a)$
		\State $s \gets$ sample according to $\Delta(last(\omega), a)$ \label{line:highprob}
		\State $\omega \gets \omega \; a \; s$
		\EndWhile
		
		\While{$\omega$ is not empty} \Comment Back-propagation
		\State $pop(\omega)$
		\State $a \gets pop(\omega)$
		\State $s \gets last(\omega)$
		\State $U(s,a) \coloneqq \sum_{s' \in S} \Delta(s,a)(s') U(s')$
		\State $L(s,a)\, \coloneqq \sum_{s' \in S} \Delta(s,a)(s') L(s')$
		\EndWhile
		\EndWhile
		\State \Return $(U(s_0), L(s_0))$
	\end{algorithmic}
\end{algorithm}

\paragraph{Variants of BRTDP}
When BRTDP is in state $s$ and it has chosen action $a$,
the choice of the next state does not necessarily have to be done by
sampling the transition distribution $\Delta(s,a)$ (as described in line \ref{line:highprob}) but can be instead chosen with probability, for instance, 
$\Delta(s,a, s') \cdot (U(s') - L(s'))$ --- which we have used in all our experiments.
One can think of other variants, for example round-robin choice.

\clearpage

\section{Explored state spaces} \label{app:states}
\begin{table}[H]
	\centering
	\caption{Number of states explored by each algorithm. For VI, the number of explored states is same as the total number of states in the model.}
	{\renewcommand{\arraystretch}{1.3}%
\begin{tabular}{lrrrrrr}
	\toprule
	Model & C & ~~~\thead{BMCTS} & ~~~\thead{MCTS-\\BRTDP} & ~~~\thead{BRTDP-\\UCB} & ~~~\thead{BRTDP} & ~~~~~~~~~~~~\thead{VI} \\
	\midrule
	coin & 4 & 4,382 & 4,411 & 6,970 & \multirow{2}{*}{7,269} & \multirow{2}{*}{22,656} \\
	& 25 & 7,229 & 7,263 & 6,912 & & \\
	leader & 4 & 140,262 & 145,526 & 137,674 & \multirow{2}{*}{138,557} & \multirow{2}{*}{237,656} \\
	& 25 & 161,619 & 137,663 & 139,381 & & \\
	mer & 4 & - & 1,013 & - & \multirow{2}{*}{3,373} & \multirow{2}{*}{17,722,564} \\
	& 25 & - & 4,710 & - & & \\
	firewire & 4 & 624 & 673 & 716 & \multirow{2}{*}{737} & \multirow{2}{*}{212,268} \\
	& 25 & 620 & 679 & 714 & & \\
	wlan & 4 & 748 & 474 & 629 & \multirow{2}{*}{541} & \multirow{2}{*}{5,007,548} \\
	& 25 & 951 & 530 & 621 & & \\
	zconf & 4 & 1,330 & 1,198 & 4,201 & \multirow{2}{*}{1,138} & \multirow{2}{*}{5,812,171} \\
	& 25 & 1,380 & 1,192 & 4,159 & & \\
	comp-firewire & 4 & 2,306 & 2,363 & - & \multirow{2}{*}{-} & \multirow{2}{*}{849,072} \\
	& 25 & 1,551 & 1,658 & - & & \\
	comp-wlan & 4 & 988 & 1,137 & - & \multirow{2}{*}{-} & \multirow{2}{*}{23,832,932} \\
	& 25 & 1,541 & 1,794 & - & & \\
	comp-zconf & 4 & - & 1,231 & - & \multirow{2}{*}{-} & \multirow{2}{*}{12,007,644} \\
	& 25 & - & 3,282 & - & & \\
	branch-firewire & 4 & 634 & 640 & 715 & \multirow{2}{*}{634} & \multirow{2}{*}{212,273} \\
	& 25 & 713 & 673 & 712 & & \\
	branch-wlan & 4 & 1,065 & 526 & 657 & \multirow{2}{*}{559} & \multirow{2}{*}{5,958,238} \\
	& 25 & 813 & 465 & 677 & & \\
	branch-zconf & 4 & 2,559 & 4,098 & 141 & \multirow{2}{*}{141} & \multirow{2}{*}{5,812,176} \\
	& 25 & 2,509 & 4,031 & 141 & & \\
	\bottomrule
\end{tabular}
	}
	\label{tab:states}
\end{table}

\end{document}